\documentclass[amsmath,amssymb,showkeys,aps,pra,reprint,footnoteinbib]{revtex4-2} 
\usepackage{amsmath,amssymb}
\usepackage[english]{babel}
\usepackage{ bbold }
\usepackage{upgreek}
\usepackage{dsfont}
\usepackage{graphicx}
\usepackage{comment}
\usepackage{braket}
\usepackage{colortbl}
\usepackage{blindtext}
\usepackage{hyperref}
\begin{document}

\title{Quantum experiments with microscale particles}

\author{James Millen}
\affiliation{Department of Physics, King's College London, Strand, London, WC2R 2LS, United Kingdom}
\email{james.millen@kcl.ac.uk}

\author{Benjamin A. Stickler}
\affiliation{Faculty of Physics, University of Duisburg-Essen, Lotharstra{\ss}e 1, 47048 Duisburg, Germany}
\affiliation{QOLS, Imperial College London, Exhibition Road, London SW72AZ, United Kingdom}

\begin{abstract}
Quantum theory is incredibly successful, explaining the microscopic world with great accuracy, from the behaviour of subatomic particles to chemical reactions to solid-state electronics. There is not a single experimental finding challenging its predictions, and ever more quantum phenomena are exploited in technology, including interferometric sensing and quantum cryptography. In order to explore novel applications and test the validity of quantum physics at the macroscale researchers strive to prepare ever heavier and bigger objects in quantum superpositions. Experiments with levitated microscale particles are about to push this quest into uncharted waters.
\end{abstract}

\maketitle

\section*{Introduction}

Despite the incredible successes of quantum theory over the last 100 years, debates continue as to its universal validity. Applying the predictions of quantum theory to macroscopic objects has always puzzled physicists and philosophers alike \cite{arndt2014a}. While decoherence theories can explain the emergence of classicality \cite{schlosshauer2019}, the fundamental question remains as to whether basic quantum phenomena, such as the archetypal superposition principle, are universally valid \cite{bassi2013}. This has been famously acknowledged by Anthony Leggett, admitting that `by day, you would see me sitting at my desk solving Schr{\"o}dinger's equation' while at night he is convinced that `at \textit{some} point between the atom and the human brain [Schr{\"o}dinger's equation] not only may but \textit{must} break down' \cite{Boer1986}. Despite this conviction shared by many, today's experiments impressively demonstrate quantum phenomena in ever larger systems: remote entanglement of picogram mechanical oscillators \cite{riedinger2018,marinkovic2018,ockeloen2018}, micrometer-spaced and seconds-lasting superposition states of atoms \cite{xu2019}, and interference of Bose-Einstein condensates \cite{berrada2013} and massive molecules \cite{Fein2019,brand2020}. These observations push modifications of quantum physics \cite{bassi2013} farther and farther into the macro-world.

Levitated optomechanics, the optical manipulation of microscale particles trapped in vacuum \cite{millen2020}, promises to push quantum experiments into an unprecedented mass regime. The field has demonstrated an exquisite degree of control over the particles, and the ability to isolate them from their environment, both key requirements for exploring quantum physics. Further striking experimental breakthroughs, including cooling massive dielectric particles deep into the quantum regime \cite{delic2020}, rotational cooling of aspherical objects \cite{delord2020,bang2020}, and ultra-precise force and torque sensing \cite{ranjit2016,kuhn2017b,ahn2020}, open the door to high mass quantum interference experiments.

In terms of the mass involved, levitated optomechanics offers to take us far beyond the state-of-the-art in generating large quantum superpositions. Nanoparticles in the quantum regime might contribute to the search for physics beyond the standard model \cite{moore2020} and future quantum technologies \cite{aspelmeyer2014}. These applications are driven by the   environmental isolation of objects levitated in ultra-high vacuum and their resulting sensitivity to external forces and torques. The ability to rotate offers great opportunities for as-of-yet unobserved quantum effects and distinguishes levitated particles from atoms and clamped oscillators.

\begin{figure}[!t]
	 {\includegraphics[width=0.48\textwidth]{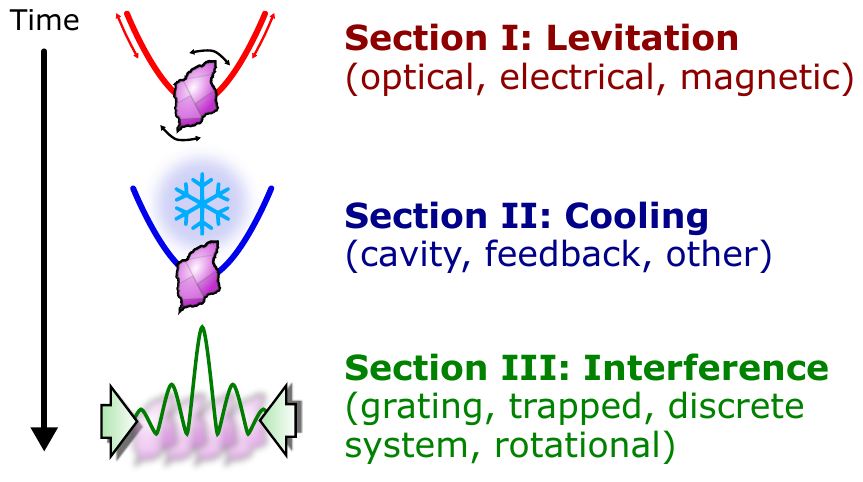}}	
\caption{\label{fig:1} 
Macroscopic quantum interference experiments require three stages of particle manipulation: levitation, cooling, and interference/detection. This article reviews strategies for each step.}
\end{figure}

In this article, we will outline strategies for producing and verifying quantum superpositions of massive objects at the quantum-to-classical borderline (see Fig.~\ref{fig:1}). We first consider how to trap, isolate and control these large particles, before introducing methods to cool their motion such that they can act as a source for an interferometer. Interferometry is the key technique for verifying the existence of a quantum superposition. We will present methods for performing interferometry, and outline the decoherence processes which may prevent the success of such experiments. We finish with a discussion of \textit{exactly what} these experiments tell us about Nature, and their implications for other areas of science and technology.

\section{Levitating microscale particles}

\begin{figure*}[!ht]
	 {\includegraphics[width=0.98\textwidth]{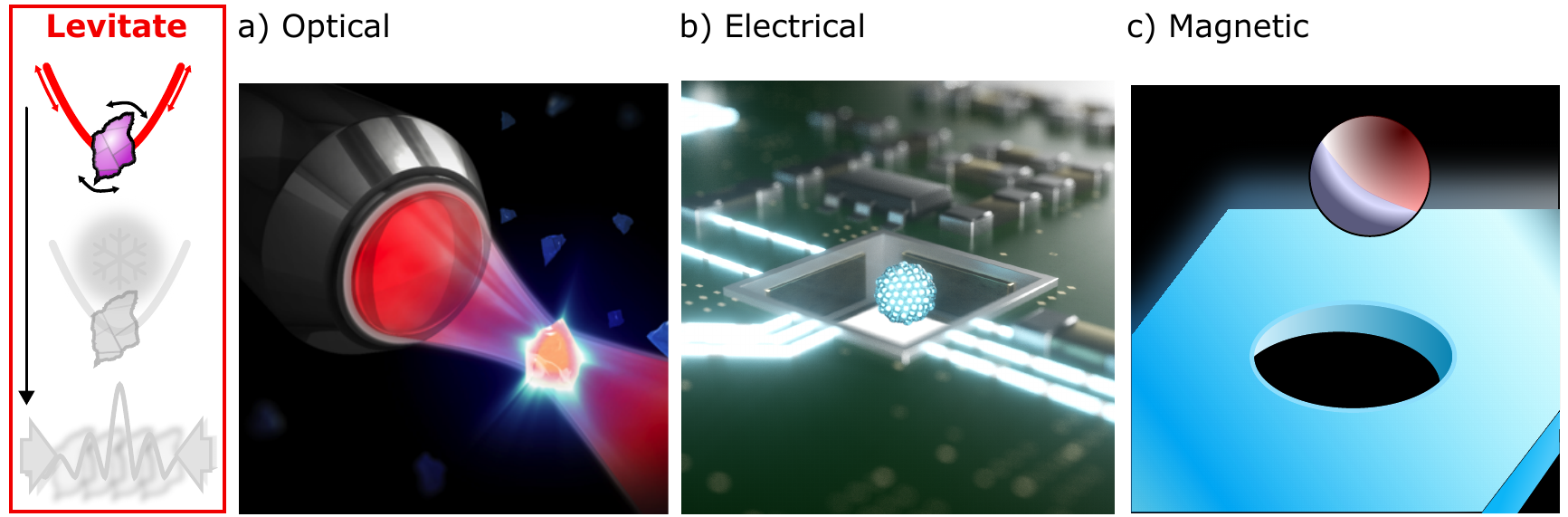}}	
\caption{\label{fig:lev} 
Microscale partices levitated in vacuum can be used for quantum experiments.  a) Optical levitation uses the gradient force exerted on dielectrics in strong laser fields, figure reproduced with permission of P. F. Barker. b) Electrical levitation uses the time-averaged potential of oscillating electric fields to suspend charged particles. c) Magnetic levitation has several variations, including the use of punctured superconducting surfaces, which expel the magnetic field of a permanent magnet.}
\end{figure*}

The first ingredient in performing quantum experiments with single particles is the source. This in itself is a two-stage process; a micro-object must be stably held in vacuum (to prevent heating from the environment), and its motional energy must be reduced.

Here, we briefly review optical, electrical, and magnetic strategies for levitating nano- and micro-particles under vacuum conditions, Fig.\,\ref{fig:lev}. In general, it is important to have control over all external degrees-of-freedom of the particle (i.e. translation and rotation), since they will interact with their environment, external fields, and each other in a complex way.  

\subsection{Optical levitation}

In the 1970's, Arthur Ashkin developed a technique for using light to confine microscopic objects, for which he was jointly awarded the Nobel Prize in 2018. This technique is known as optical tweezing \cite{Tweezer2020}, and is an established technology commonly utilized in soft-matter and medical physics. A focussed and linearly polarized optical field exerts a conservative gradient force and torque on the particle, as well as a non-conservative radiation pressure force, see Fig.~\ref{fig:lev}a). If the particle is smaller than the wavelength, this provides a stable trapping potential at the position of maximum laser intensity, where the particle's axis of maximum susceptibility aligns with the local field polarization.

For a focussed linearly-polarized Gaussian laser beam, as long as the amplitude of motion remains much smaller than the laser wavelength, the restoring force is linear, yielding independent harmonic oscillation frequencies for the centre-of-mass motion and alignment. Expressions for these frequencies can be found elsewhere \cite{Tweezer2020, millen2020}. The optical trap could be a single tightly focussed beam, an optical standing wave, or the field inside an optical cavity. 

Optical levitation is suitable for dielectric particles with radii larger than a few nm. Optical traps exhibit instability in vacuum due to relatively low trap depths \cite{millen2014}, requiring cooling in these conditions. Metallic particles can be trapped via plasmonic forces \cite{jauffred2015}, but this leads to extreme heating and hence isn't possible at pressures lower than ambient. 

In vacuum, the transfer of momentum from photons to the particle, known as photon shot noise \cite{jain2016}, and optical absorption leads to the particle heating up. This cannot be sufficiently minimized as in other quantum optical experiments, due to the strong fields required for levitation. Nonetheless, optical tweezing remains the state-of-the-art levitation technology, successfully deployed for cooling to the quantum regime \cite{delic2020}.

\subsection{Electrical levitation}

Heating caused by optical photons has motivated researchers to explore the use of other fields to harmonically confine microscale particles. The first alternative technique we consider is the ion trap, which uses electrical fields to levitate charged particles, see Fig.~\ref{fig:lev}b). This mature technology is mainly used to confine atomic ions for quantum science applications, or in mass spectrometers, but ion traps are capable of suspending any charged object, ranging in size from the atomic up to a few hundred micrometres, only limited by requirements on the electric field. 

In an ion trap, an oscillatory time varying electrical field is used to generate a harmonic pseudo- (time averaged) potential for charged particles, with the trapping frequency proportional to the charge to mass ratio. For a quadrupole field, the particle motion depends upon the interaction of its dipole and quadrupole moments with the local field \cite{martinetz2020}. The electric dipole vector tends to align with the local field direction, inducing rotational-translational coupling, whereas the quadrupole tensor tends to align its axis with the principal trap axes \cite{delord2017b,martinetz2020}. Since homogeneously charged aspherical particles have strong quadrupole moments despite having small dipole moments \cite{delord2017b,martinetz2020}, their rotational and translational motion approximately decouple in a quadrupole ion trap.

The advantages of using ion traps include extremely deep potentials allowing confinement in ultrahigh vacuum, flexibility in the objects which can be trapped, and low noise operation. The main disadvantage is the typically low harmonic frequencies (below 1 kHz), which determine the spacing between energy states in the potential, making it hard to both reach and stay-in the ground state when the frequency is low. Charged particles are susceptible to stray electrical fields and field gradients, though the detrimental influence of electrode surface noise is small due to the large distances between electrodes and trapping region \cite{Goldwater2019}.

\subsection{Magnetic levitation}

An even lower noise method is to use magnetic levitation techniques. This can either involve diamagnetic particles levitated in inhomogeneous magnetic fields \cite{hsu2016,slezak2018,obrian2019} or magnetic materials levitated above superconducting surfaces \cite{pratcamps2017,timberlake2019,wang2019,vinante2020}, see Fig.~\ref{fig:lev}c).

Diamagnetic particles can be stably trapped in magnetic quadrupole fields. The external magnetic field induces an internal magnetization field, which is proportional but opposed to the the local field strength, so that the particle is attracted to the trap centre. The particle alignment is determined by its magnetic susceptibility tensor and the local field configuration. Such setups can also be used to levitate superfluid Helium droplets \cite{childress2017} or superconducting particles \cite{hofer2019}, providing a first step towards quantum experiments with superconducting spheres \cite{romeroisart2012, cirio2012, pino2018}.

Superconducting surfaces expel the magnetic fields of nearby permanent magnets (Meisner effect), thereby exerting a repulsive force \cite{pratcamps2017,gieseler2020,wang2019,timberlake2019,vinante2020}.
This force can be used for stable trapping by either puncturing the superconducting surface \cite{pratcamps2017} (as illustrated in Fig.~\ref{fig:lev}c)), or by freezing-in the image dipole by cooling the surface below its superconducting transition in the presence of the particle. Levitated magnets provide a great platform for exploring rotational dynamics \cite{rusconi2016,rusconi2017a,rusconi2017b} and quantum magnonics \cite{gonzalezballestero2020a,gonzalezballestero2020b}.

Typical oscillation frequencies are low, below 100\, Hz \cite{pratcamps2017, slezak2018}, but can reach the kHz level \cite{gieseler2020b}. There are many advantages of magnetic levitation, including natural synergy with cryogenic operation (though this increases experimental complexity) and the ability to work with particles from 100\,nm to mm scale.

\section{Quantum cooling}

\begin{figure*}[!ht]
	 {\includegraphics[width=0.98\textwidth]{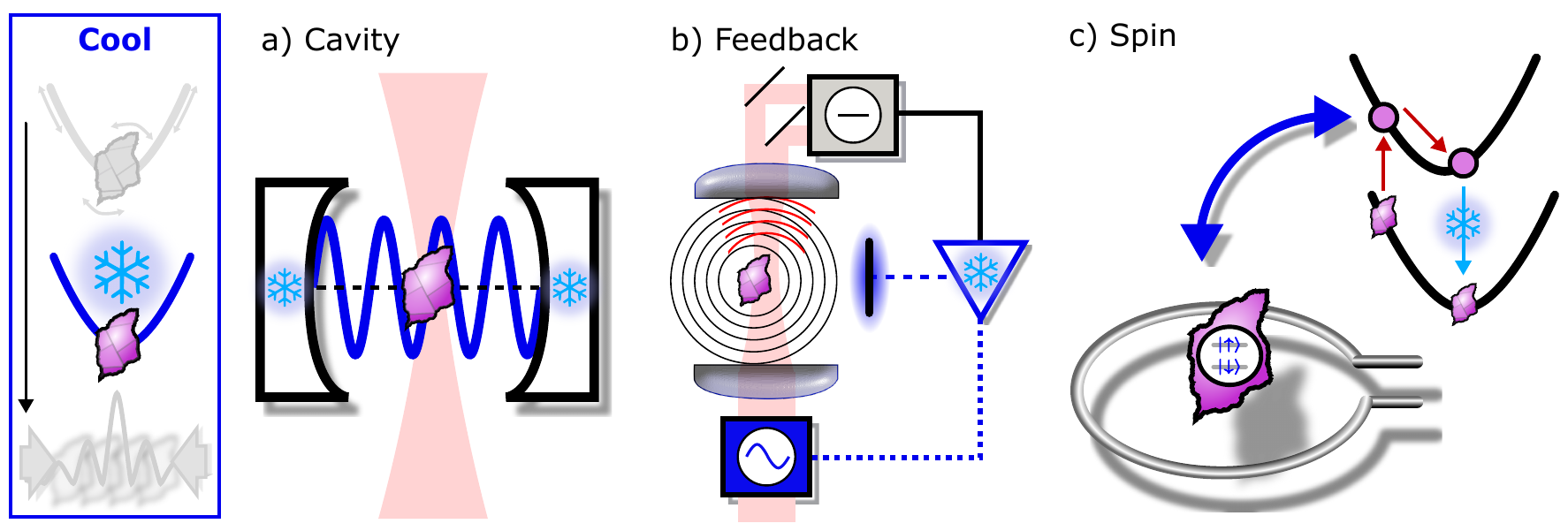}}	
\caption{\label{fig:cool} 
To observe quantum effects, the motion of microscale particles must first be cooled. a) Optical cooling via the mode of a high finesse optical cavity. b) Active feedback cooling based on continuous measurement of the particle's motion. Parametric feedback can be applied directly onto the trapping field, or cold-damping feedback can be exerted via an external force. c) Spin-cooling via driving an internal transition that couples to the particle motion.}
\end{figure*}

Now that we have levitated the particle we will use for quantum experiments, we have to cool its motion, Fig. \ref{fig:cool}. When the motion of a particle is `hot', it occupies a broad distribution of states within the trapping potential. Subsequent evolution will average over all of these states, removing quantum signatures such as the interference pattern which is evidence of a quantum superposition. Cooling also leads to greater localization, ensuring that the particle remains in the linear region of the potential where each degree-of-freedom can be independently controlled.

\subsection{Optical cavity cooling}

Optical cavity cooling, or cavity-assisted sideband cooling, is a key technique in the field of cavity optomechanics \cite{aspelmeyer2014}, see Fig.\,\ref{fig:cool}a). It relies upon the mechanical modulation of the optical path length of the cavity, allowing coherent exchange of energy between the optical field and the mechanical motion. This modulation can arise from the physical motion of one of the mirrors in a Fabry-Perot cavity, the breathing motion of a whispering gallery mode or photonic crystal resonator, or the physical motion of a dielectric levitated within the field of the cavity. 

Explicitly considering the latter case, cavity cooling relies on two processes. Firstly, the motion of the levitated particle produces sidebands on light incident upon the particle. The higher frequency `anti-Stokes' sideband corresponds to the particle losing energy through scattering, and the lower frequency `Stokes' sideband to the particle gaining energy. Secondly, the optical cavity dissipates energy through its mirrors, as quantified by its linewidth. 

The anti-Stokes sideband can be tuned into resonance with the optical cavity, with a corresponding suppression of the Stokes sideband, enhancing the loss of energy by the particle to the optical field. This corresponds to passive cooling of the particle's motion. 

There are three setups for performing optical cavity cooling, see Fig.~\ref{fig:cool}a). The optical cavity field can be used for levitation and cooling \cite{barker2010,chang2010,romeroisart2010}, a separate field can be used for levitation while the optical cavity is pumped with light to produce cooling \cite{millen2015}, or the scattering from a separate levitating optical field can be used to directly pump the cavity \cite{delic2019,windey2019,gonzalezballestero2019}. The latter configuration, known as \textit{coherent scattering}, has enabled the amazing feat of ground state cooling of a silica nanoparticle of diameter $\sim140$\,nm \cite{delic2020}. Cavity cooling works on both the centre-of-mass and alignment degrees-of-freedom \cite{stickler2016a, schafer2020}.

There are some challenges in working with optical cavities, such as delicate alignment and stabilization requirements. However, the advantages of cavity cooling include the proven ability to cool to the ground state, and the ability to produce strong light-matter coupling \cite{sommer2020}.

\subsection{Feedback cooling}

It is possible to take a more active approach to cooling, by tracking the motion of the particle and using this to generate a feedback force, see Fig.\,\ref{fig:cool}b). If this force is proportional to the velocity of the particle, it acts as a viscous damping, cooling the particle. For brevity, we will summarize the different feedback forces, without detailing the various methods by which the position signal can be acquired \cite{millen2020}.

For large particles ($>1\,\mu$m), the scattering force exerted by separate laser beams can be used for cooling. Such cooling has been demonstrated in optical \cite{li2011}, magnetic \cite{slezak2018} and ion \cite{dania2020} traps, and enables cooling to a few mK.

For smaller particles, the radiation pressure force becomes negligible. The trapping potential itself can be amplitude modulated in a technique known as parametric feedback cooling \cite{gieseler2012}, see Fig.\ref{fig:cool}b). Since the force acting on a trapped particle is symmetric about the centre of the trap, this method requires a feedback signal oscillating at twice the motional frequency. This technique has allowed centre-of-mass cooling to $\sim 100\,\mu$K \cite{jain2016}, and has been used to cool the alignment degrees-of-freedom of a non-spherical particle \cite{bang2020}. The limitation of parametric feedback cooling is that the feedback signal and the detected signal are not independent, and so the scheme requires a variable feedback gain to be efficient.

Finally, it is possible to use electrical or magnetic fields to exert a feedback force, in a process known as cold damping, see Fig.~\ref{fig:cool}b). Feedback via electrical fields onto optically trapped charged particles has been very successful \cite{tebbenjohanns2019, conangla2019}, cooling to the point at which sideband-asymmetry becomes evident \cite{tebbenjohanns2020}. Cold damping in an ion trap has also been demonstrated \cite{dania2020}, and is proposed in magnetic traps \cite{pratcamps2017}.

\subsection{Other methods of cooling}
It has been proposed to cool levitated microscale particles through coupling to fluctuating, but very cold, environments. Clouds of cold atoms are suggested as such a refrigerant for optically trapped particles, with ground-state cooling predicted \cite{ranjit2015b}. The Johnson-Nyquist noise of a cooled circuit is proposed as a refrigerant for a charged particle coupled to the circuit \cite{Goldwater2019}, and with the addition of self-sustaining feedback is predicted to allow ground-state cooling. Magnetic forces can be used to couple particles to superconducting quantum circuits, allowing deep cooling \cite{cirio2012,romeroisart2012}.

For microparticles with embedded spins, the magnetization of the spin system can be coupled to the motion of the particle, see Fig.~\ref{fig:cool}c). This has allowed cooling of the librational modes of a levitated microdiamond by driving embedded nitrogen vacancy (NV) centres \cite{delord2020}. Another type of resonance which can be exploited is that due to internal geometric optical modes such as whispering gallery modes. Proposals exist for cooling the centre-of-mass of levitated dielectric spheres \cite{barker2010b} or superfluid droplets \cite{childress2017} via this method.

\section{Quantum interference and detection} \label{sec:int}
\begin{figure*}[!ht]
	 {\includegraphics[width=0.98\textwidth]{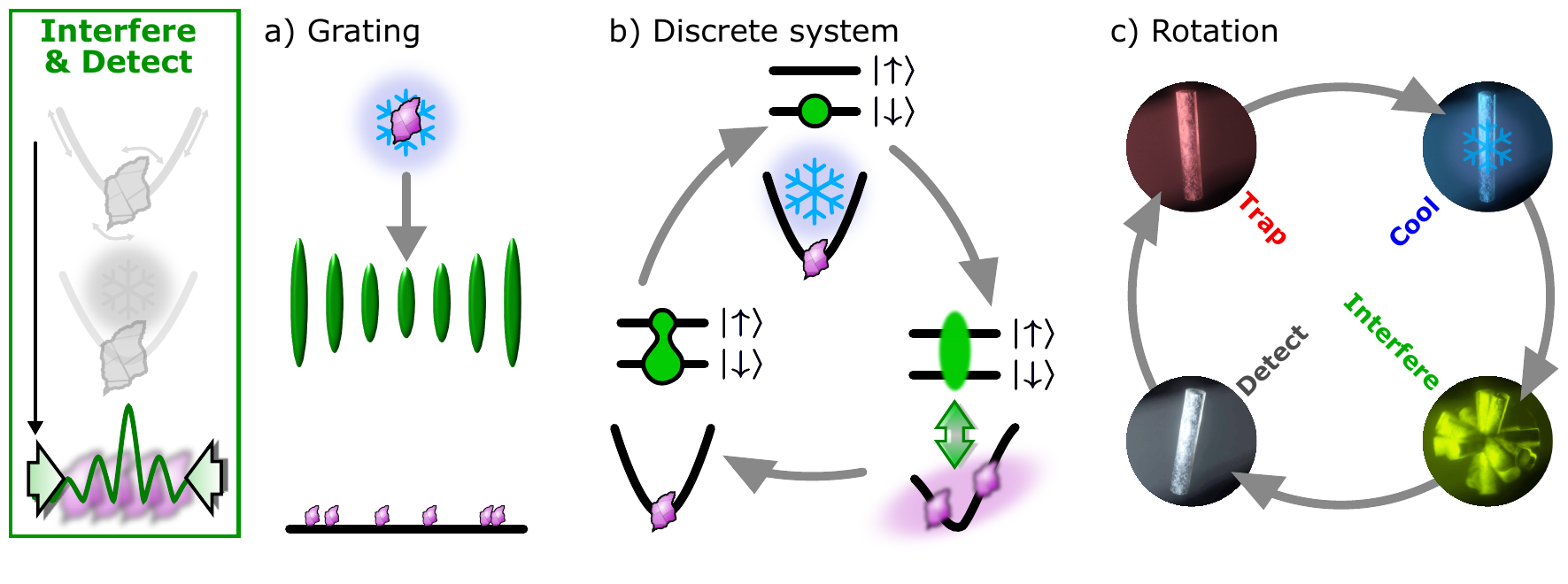}}	
\caption{\label{fig:inter} 
Quantum interference proposals can be categorized according to how the quantum superposition is generated and observed. a) Diffraction of the dispersed centre-of-mass wavefunction at an optical grating. b) Entanglement of the particle motion with auxiliary quantum degrees of freedom. c) Exploitation of the non-harmonicity of rigid-body rotations for beamsplitter-free rotational interference.}
\end{figure*}

Proposals for generating and observing motional quantum superpositions of microscale particles can be categorized according to how quantum interference is generated. Existing proposals for microscale particles use either optical gratings, coherent interaction with optical pulses, entanglement with few-level systems, or the non-linearity of rotations to achieve pronounced quantum signatures of their mechanical motion. In the following we will review the underlying concepts and briefly discuss how the interference is detected.

\subsection{Optical gratings}
Standing wave gratings are well established in atomic and molecular matter-wave interferometry \cite{cronin2009}, with recent highlights including the first demonstration of Bragg diffraction of organic molecules through thick laser beams \cite{brand2020} and the record-breaking near-field Talbot-Lau interference of $2.5\times 10^4$\, amu molecules \cite{Fein2019}. The particles start from a point source and freely disperse into a superposition of different centre-of-mass positions before hitting the grating. Through the coherent interaction with the laser grating, the delocalized particle state acquires a position-dependent phase proportional to the local field intensity. This periodic phase modulation transfers a superposition of momentum kicks, thus leading to quantum interference behind the grating if multiple grating slits were coherently illuminated. In the near field, close to the grating, different diffraction orders are not yet spatially separated and the quantum state at the grating completely recurs at integer multiples of the quantum characteristic {\it Talbot length} \cite{hornberger2011,hornberger2012}.

Near-field interference with nanoscale particles provides a promising route towards quantum superposition in an unexplored mass regime, as demonstrated in the proposal by Bateman et al. \cite{bateman2014}, see Fig.~\ref{fig:inter}a). The proposed experiment consists of first cooling the motion of a single $10^6$\,amu nanoparticle in an optical tweezer to milli-Kelvin temperatures, then releasing it so that its centre-of-mass wavefunction disperses, then briefly illuminating the particle with a standing laser wave to implement the diffraction grating, and finally, after freely evolving for a pre-determined time, adsorbing the particle on a glass plate. The collected particles will form a characteristic near-field interference pattern on the glass plate after many repetitions, which can be observed with optical microscopy and which must be compared with the classically expected shadow pattern. For certain evolution times and grating parameters, the density pattern contains pronounced quantum signatures and thus constitutes a test of the quantum superposition principle \cite{nimmrichter2011b}.

An alternative by Romero-Isart et al. shows how far-field double-slit interference of $10^7$\,amu particles could be observed by diffraction from a continuously monitored optical cavity mode \cite{romeroisart2011a,romeroisart2011c}. Again, individual particles are first trapped in an optical tweezer and cooled close to their motional ground state. Releasing the particle leads to dispersion, until they traverse a weakly driven optical cavity mode, whose output phase quadrature is continuously monitored in a homodyne measurement. This observes the squared-transverse position of the particle as it crosses the cavity mode, thereby implementing an effective double slit whose slit width is determined by the measurement outcome. Collecting the particles far away from the grating and post-selecting events with identical effective slit width gives rise to a far-field interference pattern, a clear signature of quantum superpositions. Realising a double-slit by continuous measurement has also been proposed for interfering ultra-massive superconducting spheres  ($\gtrsim 10^{13}$\,amu) by magneto-mechanical coupling to a superconducting cavity \cite{pino2018,romeroisart2017}.

Both proposals are realistic candidates for near-future quantum tests, despite all currently unsolved challenges, including the required uniformity in particle size and shape, maintaining the relative stability of all optical elements from shot to shot, and environmental decoherence due to background gas, black body radiation, laser scattering, and thermal emission (see Sec.~\ref{sec:deco}). As in molecular interferometry \cite{hornberger2012}, the nanoparticle rotations are not detrimental since the acquired phase only weakly depends on the orientation degrees of freedom \cite{Stickler2015b}. Moving to micro-gravity environments and recycling the particle could enable interference of even heavier objects \cite{kaltenbaek2016}, where the ultimate mass limit will be set by photon scattering decoherence during the grating transfer \cite{belenchia2019}.

\subsection{Laser-trapped nanoparticles}
Levitated cavity optomechanical systems are an ideal platform for generating light-mechanical quantum states with single photon pulses. In contrast to comparable techniques for clamped oscillators \cite{aspelmeyer2014}, these protocols can involve several of the motional and rotational degrees of freedom and they are not limited by clamping losses. Many of these protocols closely mirror the corresponding schemes in clamped optomechanics \cite{bowen2015}.

It has been proposed to generate non-Gaussian mechanical states by modulating the trapping field \cite{chang2010}, by continuously measuring the cavity output \cite{genoni2012}, by pulsed driving \cite{rakhubovsky2020,pitchford2020}, or by letting ground state-cooled particles interact with single photon pulses superimposed with the driving field \cite{romeroisart2011b}. From these schemes, modulating the trapping field has been experimentally realised with a classical thermal state \cite{rashid2016}, yielding a {\it squashed} phase space density. Combining non-Gaussian quantum states with the coherent optomechanical interaction enables implementing state transfer protocols \cite{chang2010}, entanglement between several nanoparticles \cite{chang2010}, teleportation of arbitrary optical inputs \cite{romeroisart2011b}, and eventually sensing beyond the standard quantum limit \cite{aspelmeyer2014}. Observing these quantum states requires either transferring the mechanical quantum state onto the optical mode and homodyning the cavity output \cite{chang2010} or performing mechanical state tomography via time-of-flight measurements \cite{romeroisart2011b}.

Most of the above protocols rely on the linear optomechanical Hamiltonian and thus work for both an externally driven cavity mode and the coherent scattering setup. Two recent proposals show how coherent scattering into a single cavity mode can generate entanglement between two co-trapped nanoparticles \cite{rudolph2020,chauhan2020}. 

While the coherence time of optically trapped nanoparticles is limited by unavoidable photon scattering decoherence, laser-trapped quantum experiments are attractive for sensing at the quantum limit and for testing quantum non-locality of massive objects \cite{chang2010,romeroisart2011b}, see Sec.~\ref{sec:persp}.

\subsection{Coupling to few-level systems}
Atom interferometers exploit the discrete internal levels of atoms to generate and read-out their spatial superpositions, enabling inertial sensing with unrivaled precision \cite{xu2019}. In contrast to microscopic quantum systems, nanoscale dielectrics lack such discrete internal levels that could be coherently addressed by laser pulses. However, nanoparticles can be artificially endowed with few-level systems by embedding single NV centres \cite{yin2013,scala2013,wan2016a,wan2016b,rahman2019} or by entangling them with superconducting \cite{romeroisart2012,cirio2012,johnsson2016,martinetz2020} or ion \cite{pflanzer2013,chen2017} qubits.

All these strategies aim at utilizing a two- or three-level system linearly coupled to the motion of the nanoparticle to control its quantum state, see Fig.~\ref{fig:inter}b). Motional superpositions can then be generated by first preparing e.g. the two-level system in a quantum superposition of ground and excited state, and then applying a state dependent-force on the centre of mass, thereby entangling the discrete and the motional degrees of freedom. After reversing the two-level system to close the interferometer, a relative phase due to the spatial superposition can be transformed into a qubit-population imbalance.

For nanodiamonds with embedded NV spins, this scheme can be enacted by exploiting the spin-dependent force in an inhomogeneous magnetic field \cite{yin2013,scala2013,wan2016a,wan2016b,rahman2019}. In a similar fashion, dielectric nanoparticles can be entangled with co-trapped ion qubits through a high finesse optical cavity mode \cite{pflanzer2013}, highly charged nanoparticles in ion traps can be conditionally displaced by an electric field determined by the state of a superconducting charge qubit \cite{martinetz2020}, and magnetic particles can be inductively coupled to flux qubits \cite{romeroisart2012,cirio2012,johnsson2016}.

Quantum interference protocols based on coupling the nanoparticle motion to external or embedded discrete levels offer a promising route for observing quantum superpositions of heavy objects. While some of these proposals remove the detrimental effects of photon scattering and absorption, they are all limited by the decoherence of the two- or three-levels system. Even with state-of-the-art technology, this precludes quantum superposition tests on a timescale comparable to other proposals \cite{romeroisart2011a,bateman2014,pino2018,stickler2018b}, however enables short timescale interference of heavier particles, so that these proposals could contribute to fundamental tests of quantum physics. For instance, observing the gravity mediated entanglement between two nanoparticles with embedded NV centres has been proposed for testing the quantum nature of gravity \cite{bose2017} (see Sec.~\ref{sec:persp}).

All these setups can be used to generate and observe entanglement between several nanoparticles by simultaneously enacting interference protocols on distant nanoparticle-qubit setups with initially entangled qubits. In addition, qubits can strongly affect the nanoparticle rotations \cite{cirio2012,ma2016,martinetz2020,pedernales2020a}.

\subsection{Rotational interferometry}
The non-harmonicity of free quantum rotations of rigid objects can give rise to pronounced orientational interference effects which have no analogy in free centre-of-mass motion \cite{stickler2018b,ma2020}. This opens the door for novel strategies to test the quantum superposition principle without beamsplitters \cite{stickler2018b,ma2020}, thereby removing a major technological hurdle and decoherence source.

The proposal by Stickler et al. \cite{stickler2018b} exploits that freely evolving linear rigid rotors, e.g. thin cylinders, return to their initial state after an integer multiple of the quantum characteristic {\it revival time}, see Fig.~\ref{fig:inter}c). Such revivals are independent of the initial state and the ensuing revival time is determined solely by the rotor's moment of inertia over Planck's constant. The revivals are caused by quantum interference of the discrete angular momentum states, and thus their observation tests quantum mechanics. Orientational quantum revivals are realistically observable for $50$\,nm double-walled carbon nanotubes, initially aligned with an optical tweezer and cooled to milli-Kelvin rotation temperatures. Upon release, the rotor orientation quickly disperses into a superposition of all possible orientations, before recurring at the much later revival time. By measuring the rotor orientation by light scattering, the interference signal becomes observable upon many repetitions. As for nanoparticle centre-of-mass interference, the requirements on laser alignment and particle uniformity can be significantly reduced by recycling the particle.

An alternative proposal by Ma et al. \cite{ma2020} uses that the free rotations of asymmetric rotors, subject to the tennis-racket instability, can turn persistent at high rotation speed. The required rotation speeds and temperatures are realistically achievable in the coherent-scattering setup \cite{schafer2020}.

In summary, rotational interference provides an exciting alternative to beamsplitter-based centre-of-mass proposals. While most requirements posed by environmental decoherence are identical, removing the beamsplitter renders rotational interferometry technologically attractive.

\section{Environmental decoherence} \label{sec:deco}
Decoherence refers to the gradual decay of a system's ability to show quantum interference, and is caused by its irreversible interaction with an environment \cite{hornberger2009}. Decoherence can explain the emergence of classicality in open quantum systems \cite{breuer2002}, and it poses a fundamental challenge to quantum superposition tests because perfect isolation is impossible. Common sources of environmental decoherence in quantum experiments with microscale particles include collisional decoherence with residual gas atoms, Rayleigh scattering decoherence of tweezer or cavity photons, and thermal radiation from hot nanoparticles. Depending on the chosen setup, other decoherence mechanisms can become relevant, such as scattering and absorption of black-body photons emitted from nearby surfaces, charge- and magnetization-induced decoherence due to fluctuating fields, or decoherence of entangled few-level systems.

It is instructive to consider decoherence due to individual uncorrelated scattering events with environmental particles, such as gas atoms or photons \cite{hornberger2008}. As a rule of thumb, interference is lost if the system reveals {\it which-way information} through the scattering process, i.e. if an idealised measurement of the environmental particle after the collision would allow distinguishing between the different pathways contributing to the interference. For instance, the quantum state of a nanoparticle delocalized over several standing laser-wave maxima can collapse to a localized state due to Rayleigh scattering of a single photon, while decoherence of a superposition of ground and first excited state in one of the laser minima requires scattering of many photons. In total, the maximally achievable coherence time for quantum superposition tests aiming at large path separations can thus be conservatively estimated as the inverse total environmental scattering rate.

For microscale particles, the rate of environmental scattering processes is typically much larger than the rate at which the particle moves. The scattered state of an environmental particle thus mainly contains information about the nanoparticle position and orientation, while it is independent of the particle linear and angular momentum. Interactions with the environment thus localize the nanoparticle quantum state in position and orientation \cite{stickler2016b,zhong2016}, which is accompanied by linear and angular momentum diffusion \cite{hornberger2008,papendell2017}. Environment-induced damping and thermalization usually only play a role when quantum coherence has already been lost \cite{breuer2002,stickler2018a}.

Collisional decoherence due to residual gas atoms can be eliminated by working in ultra-high vacuum, however, this becomes increasingly challenging for microscale objects. The rate of thermal atoms colliding with a single nanoparticle is approximately proportional to the geometric cross-sectional area of the particle and the gas pressure. Having less than a single collision during a near-field interference experiment with $10^{10}$\,amu spheres lasting $100$\,s \cite{kaltenbaek2016} thus requires stable pressures below $10^{-15}$\,mbar at $25$\,K. In addition, the cross-sectional area of non-spherical objects depends on their orientation. Thus, gas collisions also provide a relevant mechanism for decoherence in rotational interference schemes \cite{stickler2018b,ma2020}.

Photon scattering from sub-wavelength particles occurs with a rate proportional to their squared volume. This determines the achievable coherence time of optically trapped quantum experiments \cite{romeroisart2011b} and strongly limits the applicability of laser gratings for high-mass interferometry \cite{bateman2014,belenchia2019}. Absorption of laser photons leads to internal heating of the nanoparticle, so that it starts thermally radiating. While the exact form of the spectral emissivity will depend on how Planck's law is generalized for small objects \cite{rubiolopez2018}, the total rate of thermal radiation strongly contributes to decoherence \cite{bateman2014}. Even if laser gratings are avoided, scattering and absorption of black-body photons emitted from the experimental aparatus can lead to decoherence \cite{romeroisart2011b}. Eliminating photon scattering and absorption is one of the main motivations for magnetic and electric trapping schemes. While magnetization- and charge-induced decoherence mechanisms are largely unexplored, several factors could play a role, including patch charges on nearby surfaces \cite{knobloch2016} or magnetic field fluctuations \cite{pino2018}. In addition, rotational dephasing \cite{martinetz2020,pedernales2020a} and qubit decoherence seem to be very important for quantum experiments with entangled few-level systems \cite{scala2013,martinetz2020}.

\section{Perspectives}
\label{sec:persp}
With the recent breakthrough of reaching the centre-of-mass groundstate \cite{delic2020} and the first demonstrations of rotational cooling \cite{delord2020,bang2020}, further ground-breaking experiments are imminent, opening the door for interferometric tests of collapse models, exquisite force and torque sensing at the quantum limit, and novel technological applications. In the following we give a brief overview over possible future experiments and their implications.

\subsection{Macroscopic superpositions}
Quantum theory assumes the universal validity of the quantum superposition principle, implying that it should be possible to prepare quantum states where objects of arbitrary size, mass, and complexity are simultaneously at different locations in space. The most well-known alternatives to standard quantum theory are the gravitational collapse model of Di{\' o}si and Penrose \cite{diosi1989} and the model of \textit{continuous spontaneous localization} (CSL) \cite{bassi2003,smirne2015,carlesso2018b}. These models predict {\it observable} deviations from standard quantum theory, giving rise to wavefunction localization in position space at the macro-scale while being fully compatible with all experimental observations to date. Importantly, these models can be falsified in the lab \cite{nimmrichter2011b,romeroisart2011c,bahrami2014,nimmrichter2014,goldwater2016,li2016, toros2017,schrinski2017}.

The model of CSL predicts that all massive objects experience random momentum kicks from spontaneous wavefunction collapse events \cite{bassi2013}. These kicks lead to spatial decoherence as well as to motional heating as characterised by two free parameters of the theory, the collapse rate and its length-scale. The absence of collapse-induced decoherence and heating of well-isolated objects thus rules out parameter combinations implying stronger effects than observed. Heating experiments currently provide the most stringent bounds on CSL \cite{vinante2020b}, with interferometric tests following closely \cite{nimmrichter2011b,nimmrichter2013,schrinski2019,schrinski2020a,schrinski2020b}. The extent to which a given quantum superposition principle rules out objective modifications of quantum theory also provides an experimentally useful notion of {\it macroscopicity} \cite{nimmrichter2013,schrinski2019}.

Quantum interference experiments with massive objects also provide a platform for falsifying and even sensing dark matter-induced decoherence \cite{riedel2013}. While collisions with low-mass dark matter have negligible classical impact on the dynamics of a molecule or nanoparticle, the scattering events can induce decoherence by extracting which way information. Dark matter-induced decoherence can be identified by observing its variation with the rotation of the Earth \cite{riedel2017}.

In addition, it was proposed that witnessing gravity-induced entanglement between two microscale particles reveals if their gravitational interaction constitutes a quantum channel \cite{bose2017,marletto2017}. Bose et al. \cite{bose2017} propose to prepare two microscale particles in a quantum superposition of two different relative distances. If gravity is quantum, the gravitational interaction entangles the motional quantum states, which can be transferred to and witnessed by embedded NV centre spins \cite{bose2017,marletto2017}. The proposed masses, coherence times, and experimental requirements are many orders of magnitude beyond what is currently achievable \cite{vandekamp2020,chevalier2020}. Nevertheless, the prospect of lab-based tests of quantum gravity is accelerating the development of cutting edge microparticle manipulation techniques.

\subsection{Other fundamental physics}

The force and torque sensitivity of levitated nanoparticles in the deep quantum regime surpasses that of state-of-the-art devices by several orders of magnitude \cite{millen2020,moore2020}, without having to be prepared in macroscopic superposition states. Levitated nanoparticle sensors are ideally suited for thorough studies of the electrostatic and electrodynamic forces and torques exerted by nearby surfaces \cite{xu2017,winstone2018,ahn2020,kawasaki2020} or electric circuits \cite{Goldwater2019,braidotti2020}. For instance, particles in front of a bi-refringent plate experience a Casimir torque due to vacuum fluctuations of the electromagnetic field \cite{parsegian1972}. This torque has only recently been measured \cite{somers2018}, and levitated nanoparticles provide an ideal platform for exploring it at ultra-low surface and particle temperatures \cite{xu2017}. If the particle is moving or rotating, the Casimir force and torque can slightly lag behind due to the finite conductivity of the surface. The resulting drag is referred to as {\it vacuum friction} \cite{zhao2012} and will become observable with levitated nanorotors close to surfaces \cite{ahn2020}.

Levitated nanoparticles are also well suited for explorations of gravity at ultrashort distances and for testing physics beyond the standard model. How these systems can probe deviations from Newton's gravitational force law at short distances \cite{geraci2015,rider2016}, for detecting high-frequency gravitational waves \cite{arvanitaki2013}, or for excluding the existence of milli-charged particles \cite{moore2014} has been reviewed recently \cite{moore2020,carney2020}.

\subsection{Technological applications}

The prospect of using levitated particles as state-of-the-art sensors in the context of exploring fundamental physics is discussed above and elsewhere \cite{millen2020,moore2020}. In terms of industrial applications, inertial sensing is key for the automotive, aerospace and medical sectors, and gravimetry is regularly deployed for oil and gas prospecting. The current cutting-edge sensors are based on atom interferometry, but due to their increased mass, interference of microparticles offers five orders of magnitude improvement in sensitivity \cite{qvarfort2018,rademacher2020}.

In terms of the detection of other forces, levitated particles are at the extreme low-mass end of nanomechanical sensors, making them sensitive to the smallest forces \cite{ranjit2016}. The geometry offers other advantages compared to clamped sensors, such as potential operation in free-fall enabling static-force detection \cite{hebestreit2018}, and their physically small size enabling the detection of forces extremely close to surfaces \cite{rider2016} (see also above). 

Other optomechanical systems already feature in emerging quantum technologies, where mechanical modes are used for coherent signal storage and transduction. When compared to other nanomechanical devices, the advantages of using levitated systems center around the high degree of control over the Hamiltonian. It is possible to vary the trapping frequency in time, allowing the generation of squeezed states \cite{rashid2016}. Although state-of-the-art clamped systems demonstrate as low-dissipation as promised by levitated objects, cryogenic environments are not necessarily required to preserve the coherence of levitated or free microparticles. When considering cavity optomechanical systems, the levitated platform has demonstrated straightforward access to quadratic, or non-linear, optomechanical coupling \cite{fonseca2016}, a key ingredient in quantum state engineering.  

Finally, we note that recent proposals for coupling the motion of levitated particles to quantum systems \cite{pratcamps2017,gieseler2020b,martinetz2020} may enable quantum bath engineering, and hence the study of single-particle quantum thermodynamics \cite{millen2016}.

\section{Outlook and conclusions}
`Measuring instruments, if they are to serve as such, cannot be included in the range of applicability of quantum mechanics.' So said Niels Bohr to Erwin Schr{\" o}dinger in 1935 \cite{Moore1992}, separating out the microscopic quantum realm and the tools we use to probe it. This admission that a fundamental theory may not be universal was surprising, but also of little practical consequence. Quantum mechanics was thought to only apply to the tiniest constituents of nature, and the assignment of a wavefunction to `macroscopic' objects impossible (Bohr would define macroscopic objects as those we can clearly observe to have fixed classical properties, and did not imagine the direct observation of, say, the motion of a molecule \cite{Whitaker2006}).

Today, we have reached a level of scientific sophistication where this question \textit{does} have practical consequence, since the realms of nanotechnology and quantum technology are beginning to overlap. In addition, our understanding of biological processes at the nanoscale is such that we are forced to ask questions about the role of quantum theory in increasingly complex systems.

Interferometry of massive objects can verify that the quantum superposition principle holds under those particular conditions, or falsify certain models that propose modifications to standard quantum theory. The absence of interference can never falsify quantum mechanics, since we would need full confidence in our models of environmental decoherence, which are still being developed for increasingly complex systems. Of course, quantum experiments with complex objects help us understand decoherence mechanisms.

As quantum technologies advance at a dizzying pace, and our devices reach ever-further into the macroscopic realm, it is time to revisit the fundamental conceptual challenges at the heart of quantum physics. Excitingly, we now have all of the tools with which to probe the limits of this most enigmatic of physical theories. 

\section*{Acknowledgments}
The authors thank Klaus Hornberger, Ying Lia Li, Lukas Martinetz, Maryam Nikkhou, Muddassar Rashid and Daqing Wang for their valuable comments on the manuscript.

JM is supported by the European Research Council (ERC) under the European Union's Horizon 2020 research and innovation programme (Grant agreement No. 803277), and by EPSRC New Investigator Award EP/S004777/1.\\

\end{document}